\documentclass[letter,structabstract]{aa}
\usepackage{enumitem}
\usepackage{graphicx}
\usepackage{txfonts}
\usepackage{epstopdf}
\begin{document}
%
  \title{The first Herschel view of the mass-SFR link in high-z galaxies \thanks{Herschel is an ESA space observatory with science instruments provided by European-led Principal Investigator consortia and with important participation from NASA.
}}
     \titlerunning{The mass-SFR link in high-z galaxies}

   \author{G. Rodighiero
          \inst{1}
        \and A. Cimatti\inst{2}
        \and C. Gruppioni\inst{3}
        \and P. Popesso\inst{4}
        \and P. Andreani\inst{5,13}
        \and B. Altieri\inst{7}
        \and H. Aussel\inst{8}
        \and S. Berta\inst{4}
        \and A. Bongiovanni\inst{9}
        \and D. Brisbin\inst{10}
        \and A. Cava\inst{9}
        \and J. Cepa\inst{9}
        \and E. Daddi\inst{8}
        \and H. Dominguez-Sanchez\inst{3}
        \and D. Elbaz \inst{8}
        \and A. Fontana\inst{6}
        \and N. F{\"o}rster Schreiber\inst{4}
        \and A. Franceschini\inst{1}
        \and R. Genzel\inst{4}
        \and A. Grazian\inst{6}
        \and D. Lutz\inst{4}
        \and G. Magdis\inst{8}
        \and M. Magliocchetti\inst{11}
        \and B. Magnelli\inst{4}
        \and R. Maiolino\inst{6}
        \and C. Mancini\inst{12}
        \and R. Nordon\inst{4}
        \and A. M. Perez Garcia
        \and A. Poglitsch\inst{4}
        \and P. Santini\inst{6}
        \and M. Sanchez-Portal\inst{7}
        \and F. Pozzi\inst{2}
        \and L. Riguccini\inst{8}
        \and A. Saintonge\inst{4}
        \and L. Shao\inst{4}
        \and E. Sturm\inst{4}
        \and L. Tacconi\inst{4}
        \and I. Valtchanov\inst{7}
        \and M. Wetzstein\inst{4}
        \and E. Wieprecht\inst{4}
          }

\institute{\centering \vskip -10pt \small \it (See online Appendix \ref{sect:affiliations} for author affiliations) }

   \date{Received March 31, 2010  accepted ....}

 \abstract
{}
{We exploit deep observations of the GOODS-N field taken with PACS, 
the \textit{Photodetector Array Camera and Spectrometer}, on board of 
\textit{Herschel}, as part of the \textit{PACS Evolutionary Probe} guaranteed time 
(PEP), to study the link between star formation and stellar
mass in galaxies to $z\sim 2$.}
{Starting from a stellar mass -- selected sample of $\sim4500$ galaxies
with mag$_{4.5 \mu m}<23.0$ (AB), we identify $\sim350$ objects with a PACS 
detection at 100 or 160 $\mu$m and $\sim1500$ with only {\textit Spitzer} 
24 $\mu$m counterpart. Stellar masses and total IR luminosities ($L_{IR}$) 
are estimated by fitting the Spectral Energy Distributions (SEDs).
}
{Consistently with other \textit{Herschel} results, we find that $L_{IR}$ 
based only on 24 $\mu$m data is overestimated by a median factor $\sim1.8$ 
at $z\sim2$, whereas it is underestimated (with our approach) up to a 
factor $\sim1.6$ at $0.5<z<1.0$. We then exploit this calibration to 
correct $L_{IR}$ based on the MIPS/\textit{Spitzer} fluxes. These 
results clearly show how \textit{Herschel} is fundamental to constrain 
$L_{IR}$, and hence the SFR, of high redshift galaxies. Using the galaxies
detected with PACS (and/or MIPS), we investigate the existence and 
evolution of the relations between the star formation rate (SFR), 
the specific star formation rate (SSFR=SFR/mass) and the stellar mass. Moreover,
in order to avoid selection effects, we also repeat this study through 
a stacking analysis on the PACS images to fully exploit the far-IR 
information also for the {\textit {Herschel}} and {\textit {Spitzer}} 
undetected subsamples.
We find that the SSFR-mass relation steepens with redshift, being almost flat at $z<1.0$
and reaching a slope of  $\alpha=-0.50^{+0.13}_{-0.16}$ at $z\sim2$, at odds with recent works based on radio-stacking 
analysis at the same redshift. 
The mean SSFR of galaxies increases with redshift, by a factor $\sim15$ for massive 
$M>10^{11}M_{\odot}$ galaxies from $z=0$ to $z=2$, and seems to flatten at $z>1.5$  in this mass range.
Moreover, the most massive galaxies have the lowest SSFR at any $z$, implying 
that they have formed their stars earlier and more rapidly than their low mass 
counterparts ($downsizing$).

}
{}

   \maketitle

\section{Introduction}
\label{intro}
The link between galaxy stellar mass and star formation rate (SFR), and its
cosmic evolution is crucial to shed light on the processes of galaxy 
formation. The specific SFR (SSFR = SFR/mass) plays an important role
as it measures the star formation efficiency of a galaxy and the 
fraction of a galaxy mass can be converted into stars per unit time 
(see e.g. de Cunha et al. 2010). Several studies at $0<z<3$ report 
similar findings: (1) the SSFR increases with redshift at all masses; 
(2) the SSFR of massive galaxies is lower at all $z$ (e.g. Feulner et al. 2005, Erb et al. 2006, Perez-Gonzalez et al. 
2008, Damen et al. 2009, Dunne et al. 2009). However, the scatter 
and the exact slopes of these relations are still not clear. 
In particular, the dependence of SSFR on mass is one of the most debated 
open questions. On the one hand, radio-stacking analysis of galaxies  at $z=1.5-2.0$
selected in the $K$-band found a weak (or even absent)
SSFR-mass correlation, with an indication of steepening at higher 
redshift (Pannella et al. 2009, Dunne et al. 2009). On the other hand, 
other results based on UV to mid-IR SFR tracers indicate a clear
decrease of the SSFR with increasing mass (e.g. Feulner et al. 2005, 
Erb et al. 2006, Noeske et al. 2007b, Cowie \& Barger 2008). 
Some of these discrepancies could be due on one hand to the
effects of dust extinction corrections for the SFR estimates based
on UV--optical indicators, and on the other hand to the assumptions
and large extrapolations on the infrared SED shape and luminosity
adopted for the SFR estimates based on 24$\mu$m data. 

The advent of the {\it Herschel Space Observatory} (Pilbratt et al. (2010) 
finally allows us to robustly derive the total infrared (IR) luminosity 
($L_{IR}$) of galaxies,  by directly sampling the peak of the thermal emission of dusty galaxies up to $z\sim3$.
In order to place new and stringent constraints on the SSFR 
evolution, in this work we exploit the PACS Evolutionary Probe (PEP) guaranteed time
data collected in the GOODS-North field with the 
PACS (Poglitsch et al. 2010) instrument. The PEP observations are 
described in Berta et al. (2010, this Issue Appendix A). Ancillary data, both 
photometric and spectroscopic,  including UV (GALEX), optical (HST), 
near-IR (FLAMINGOS, IRAC/{\textit Spitzer}) and mid-IR (MIPS/{\textit 
Spitzer}) data, have been collected to build up a reliable 
multiwavelength catalog and photometric redshifts. 
We adopt $h = 0.7$, $\Omega_{\Lambda}=0.73$ and $\Omega_m= 0.27$. 

\section{Sample selection}
Our aim is to investigate the evolutionary link between stellar 
mass and star formation avoiding strong biases and selection effects. 
Thus, the main galaxy sample was selected at 4.5$\mu$m with IRAC (mag$_{4.5 \mu m}<23.0$, AB) in order to ensure sensitivity 
to stellar mass up to $z\sim 2-3$. At this limiting magnitude, the IRAC sample 
is $\sim$80\% flux complete (Mancini et al. 2009), it is not strongly affected by confusion (see Rodighiero et al. 2010) and it includes 4459 sources. 
On the IRAC positions we fitted PSFs to the MIPS and PACS images (however, for  detections in the PACS maps we  used only IRAC positions with a 24um detection).
Out of the 4459 IRAC sources, 1887 (351) sources have MIPS (PACS) fluxes with signal-to-noise ratio (SNR) greater than 3. 
The typical 24$\mu$m, 100$\mu$m and 160$\mu$m fluxes in  this catalog 
reach the 3-$\sigma$ limit, that is $\sim$20$\mu$Jy (Magnelli et al. 2009), 
$\sim$3mJy and $\sim$5.7mJy, respectively. With this selection we miss
only 2 PACS-detected objects with low signal-to-noise ratio (SNR$\sim3$).
About 40\%, $\sim52\%$ and $\sim70\%$ of the IRAC, MIPS and PACS subsamples have an optical spectroscopic redshift (mainly from Barger et al. 2008), respectively.

\section{Stellar masses and IR luminosities}

Following the approach of Rodighiero et al. (2007), stellar masses
were estimated by setting the redshift (photometric or spectroscopic)
of each IRAC-selected object and using the $Hyperz$ code (Bolzonella 
et al. 2000) applied to the photometric SEDs in the optical-to-5.8$\mu$m 
range. For an easier comparison with literature data, we used the 
stellar-population synthesis models of Bruzual \& Charlot (2003, 
hereafter BC03) with a Salpeter IMF, exponentially declining $\tau$ 
models for the SFR, and solar metallicity. 
We estimated the stellar mass completeness
as a function of redshift as described in Mancini et al. (2009) (see online Figure \ref{chiara}).

  \begin{figure}
   \centering
\includegraphics[width=8.cm,height=7.7cm]{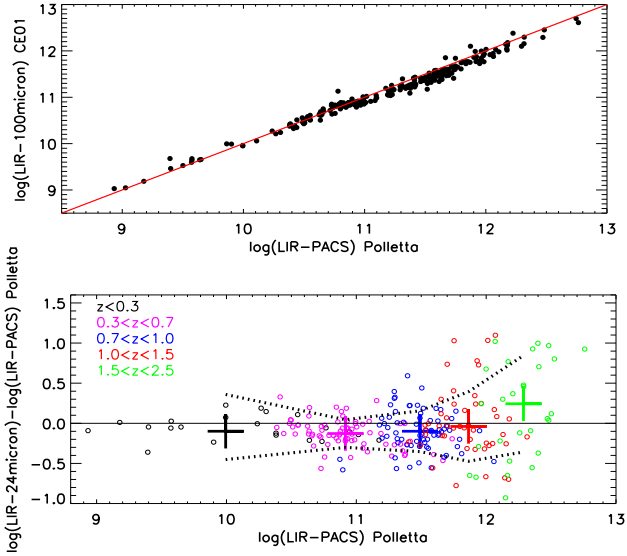}
   \caption{
     {\it Top panel}: Comparison of $L_{IR}$ derived from the UV-to-PACS$\mu$m data with the Polletta templates and the Chary \& Elbaz (2001) libraries, by fitting one single point (100$\mu$m).
      {\it Bottom panel}: Comparison of $L_{IR}$ derived from the UV-to-24$\mu$m and the UV-to-PACS SED fitting procedures. Different
   colors refer to various redshift intervals (see legend in the figure). The thick cross symbols represent the median value  in the corresponding luminosity bins. 
We also report with dotted lines the $\pm$1$\sigma$ limits around the median.
In both panels only sources from the 4.5 $\mu$m catalog that were detected with SNR$>3$ at all 24, 100 and/or 160 $\mu$m are reported.
}
             \label{lir_CE}%
    \end{figure}

 
For the objects detected with PACS and MIPS, the IR 
luminosities $L_{IR}$ were estimated with a best fitting procedure by comparing
the optical-to-IR SEDs with a library of template SEDs of local objects 
from Polletta et al. (2007) and adding a few modified templates  (see Gruppioni et al. 2010).
We followed the same approach that we applied to 24$\mu$m GOODS-South 
and SWIRE-VVDS sources in Rodighiero et al. (2010) and in Gruppioni et 
al. (2010). 
The inclusion of the whole SED in the fitting 
procedure allowed us to fully exploit the photometric information and 
determine the K-correction in the most reliable way. 
As described in Rodighiero et al. (2010),  we forced the spectral fit to reproduce the far-IR 
\textit{Spitzer-Herschel} datapoints by reducing their photometric errors. 
To minimize the contribution from AGN emission to the $L_{IR}$, we removed 
from our sample X-ray detected sources with $L_X>10^{42}$ erg/s and those objects
classified by the SED fitting analysis as Type 1 quasars.
A few examples of our best-fits are presented in the online Appendix C, Figure \ref{sedfitfig}.

\begin{figure*}[ht!]
\centering
\includegraphics[width=22.5cm,height=6cm]{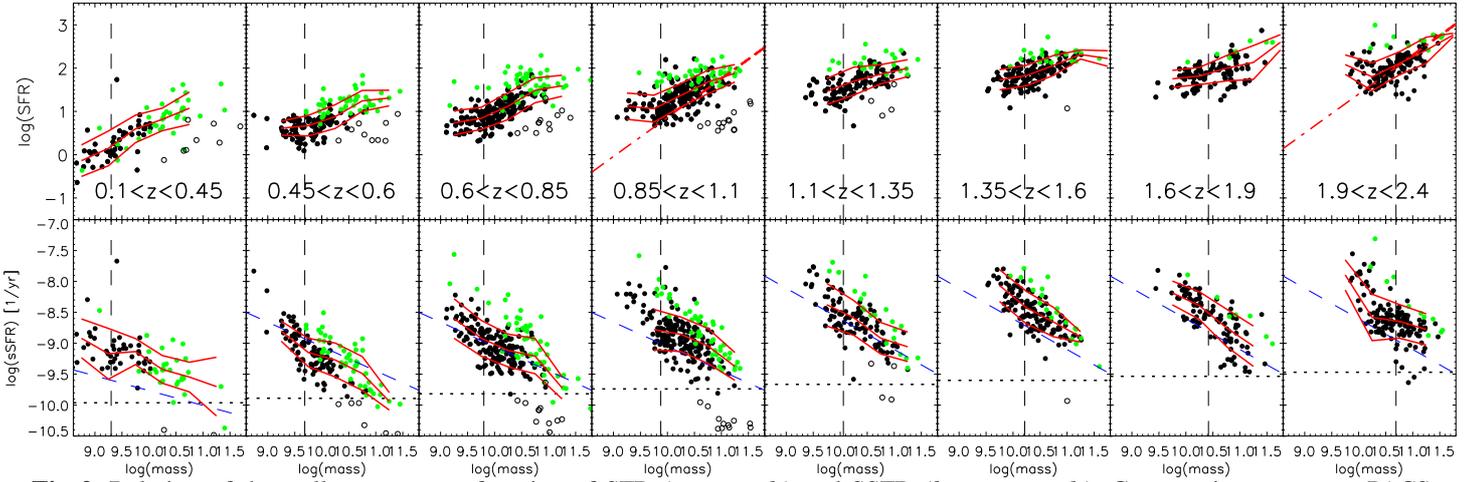}
\caption{Relation of the stellar mass as a function of SFR ({\it top panels}) 
and SSFR ({\it bottom panels}). Green points represent PACS detected sources
for which the $L_{IR}$ has been computed including all data points. Black 
points are MIPS 24$\mu$m sources undetected by PACS: in this 
case the $L_{IR}$ has been derived from our SED fitting procedure limited at 
24$\mu$m.  Open symbols are passive sources with color mag[24]-mag[3.6]$<$0.5.
We also report with red lines the average of the relations and 
the $\pm$1$\sigma$ limits. The red dot-dashed lines correspond to literature 
observed relations between mass and SFR: at $z\sim1$ the GOODS relation by 
Elbaz et al. (2007); at $z\sim2$ the relation for star-forming $BzK$ by 
Daddi et al. (2007). Vertical dashed lines mark the mass completeness limit 
as a function of $z$, and the horizontal dotted lines indicate the 
inverse of the age of the universe at the mean redshift of each bin. The 
dashed blue lines in the bottom panels show the relations derived from our 
stacking analysis on PACS maps.
Open circles mark sources classified as passive on a color basis (mag[24]-mag[3.6]$<$0.5), and generally populate  a separate sequence (this bimodality
has already been detected by Santini et al. 2009).
}
\label{mass_sfr}%
\end{figure*}

\section{The estimate of SFR}

For the galaxies detected with PACS or at 24$\mu$m, the instantaneous SFR
was estimated using the combination of IR and UV luminosity as in Papovich 
et al. (2007) and Santini et al. (2009): $SFR_{IR+UV}/M_{\odot}yr^{-1}=1.8 
\times 10^{-10} \times L_{bol}/L_{\odot}$ with $L_{bol}=2.2 \times 
L_{UV}+L_{IR}$, where $L_{IR}$ was derived by integrating the best-fit 
SEDs in the [8-1000]$\mu$m rest-frame range. The rest-frame UV luminosity 
(which accounts for for the contribution for young unobscured stars), 
uncorrected for extinction, derived from the SED fitting with BC03 model, 
is $L_{UV} = 1.5 \times L(2700 \AA)$. The contribution from $L_{UV}$ is 
marginal, and its exclusion does not change the results presented here.
In the rest of the paper, we will then discuss the main 
evolutionary trends linking  mass and SFR by considering only 
$\lambda \geq $24$\mu$m detections, and we will discuss the contribution of 
the remaining far-IR undetected IRAC sources through a stacking analysis. 

\section{Results}
\subsection{{\textit Spitzer} vs {\textit Herschel} IR luminosities}

The reliability of our IR luminosities was assessed by comparing them 
with alternative approaches. For PACS detections, we compare in Figure 
\ref{lir_CE} (top panel) our $L_{IR}$ derived from SED fitting to the 
UV-PACS range with Polletta templates, with those computed by using the 
Chary \& Elbaz (2001, hereafter CE01) library and fitting only the 
100$\mu$m fluxes. This shows a very tight correspondence, confirming 
the value of far-IR information to constrain the bolometric energy budget 
of high-$z$ galaxies, independently of the models assumed.
For sources with 24$\mu$m detection and no PACS, we have to rely on 
large extrapolations, that can be strongly influenced by the adopted models.
As an internal check, we then compare in Figure \ref{lir_CE} (bottom panel) 
the IR luminosity derived from our UV-to-24$\mu$m and the UV-to-160$\mu$m 
SED fitting procedures. Different colors refer to various redshift intervals.  
As already discussed by Elbaz et al. 2010 and Nordon et al. 2010, at $z>1.5$  
$L_{IR}$ based on 24$\mu$m data are overestimated, by a median factor 
$\sim1.8$ in our analysis: this factor is larger for the CE01 models and when including
PACS upper limits in the analysis (e.g. up to a factor $\sim4$, see Nordon et al. 2010, showing that the observed trend is not induced by any PACS incompleteness). 
At $z<1.5$, the 24$\mu$m based $L_{IR}$ are underestimated, up to a factor $\sim1.6$ 
around $z\sim0.5-1.0$. This result highlights the key role of \textit{Herschel} and have strong implications on previous {\textit {Spitzer}} 
works on the SFR of $z\sim2$ galaxies. In the following analysis and figures we have 
corrected the 24$\mu$m based $L_{IR}$, according to the new \textit{Herschel} calibration that we have just discussed.

\subsection{The mass versus SFR relation of star-forming galaxies}

The existence of a strong correlation between galaxy SFR and mass at 
different redshifts ($0<z<3$)  has been extensively discussed in the recent 
literature (Erb et al. 2006, Noeske et al. 2007a, Elbaz et al. 2007, Daddi
et al. 2007, Perez-Gonzalez et al. 2008, Pannella et al. 2009, among the others), but
there is not yet agreement on the general validity and properties of
this relation. 
The main results of our analysis are shown in Figure \ref{mass_sfr} 
where we combine the subsamples of the \textit{star-forming} galaxies 
detected with PACS and/or MIPS. 
\begin{figure}
\centering
\includegraphics[width=6.2cm]{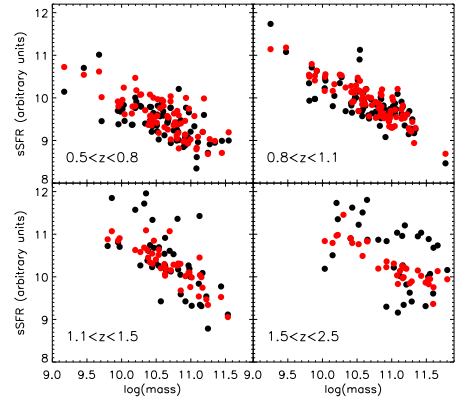}
\caption{Relation of the stellar mass as a function of the SSFR for 
PACS detected sources in various redshift bins: the SFR for red points 
has been computed from PACS fluxes, while for black points it has been 
extrapolated from the SED fitting from the 24$\mu$m data. 
}
\label{sel_effect}%
\end{figure}
The upper panels indicate the
existence of a (rather scattered) positive correlations between the 
SFR and stellar mass at all redshifts. The Spearman test indicates that 
the probability of not having correlation is 
2e-23,  7e-23,  9e-41, 1e-28,  2e-14,  2e-17,  4e-08, and 1e-10 in the eight redshift intervals, respectively. 
The SFR-mass  relation looks more robust at $z<1$. However we should remind that in 
this case we are considering also sources below the mass completeness 
(vertical lines), and that we are sampling only the most star-forming 
population, as MIPS or PACS undetected sources are not included in these 
plots. The comparison with Elbaz et al. (2007) at $z\sim1$ and Daddi et 
al. (2007) at $z\sim2$ (see caption of Figure \ref{mass_sfr} for details) 
shows that their observed slope of the SFR-mass relation 
is not inconsistent with our results, but even much flatter relations would be allowed by our data above the completeness limit.
The lower panels of Figure \ref{mass_sfr} show the relation between
SSFR and stellar mass for the 
same sources plotted in the upper panels. A negative trend of SSFR
with mass is evident at all redshifts, although the scatter is quite large.
The bulk of  PACS and/or MIPS sources is located above the 
horizontal dotted line (the inverse of the age of the Universe),
indicating that these systems are experiencing a major episode of star 
formation, forming stars more actively than in their recent past and 
building up a substantial fraction of their final stellar mass.

The key role of \textit{Herschel} for high-$z$ galaxies is highlighted 
in Figure \ref{sel_effect} showing that when PACS data are used, the 
scatter of the SSFR - mass relations decreases by factors of 1.0,
1.3, 2.3 and 3.7 for increasing redshift bins, compared to the case
of SSFRs estimated with MIPS data only.
We have to warn that PACS detects only the brightest objects and we cannot than verify that the scatter is intrinsically lower.
However, the fact that at least at high luminosities, at $z\sim2$, PACS produces a smaller scatter (because it provides a more accurate SFR),
might suggest that a similar trend  should happen also at low luminosities.



\section{Stacking analysis on PACS images}
To test if the results in Figure 2 might have been affected by selection biases and to verify their reliability, 
we performed a stacking analysis (Bethermin et al. 2010)  including all sources of the original IRAC [4.5]$<$23.0 sample. 
We splitted the sample in bins of mass and redshift, and stacked on  a residual  160$\mu$m map
(created by removing all PACS 160$\mu$m detections included in our catalog with SNR$>3$)
at the positions of all IRAC sources undetected by PACS (stacking at 
100$\mu$m does not change our results). With this procedure, we derived a
typical flux for each mass-redshift bin. Using the formalism introduced by Magnelli et al. (2009), that accounts both for detections and no-detections, 
we then converted these fluxes into luminosities by adopting an average
$K$-correction for each redshift (derived from the more common SED template 
adopted by our best-fitting procedure in that interval). An empirical 
relation between rest-frame $\nu L_{\nu}(160\mu m)$ and $L_{IR}$ was used 
to first convert the stacked luminosities into bolometric luminosities, and 
then into SFR through a standard Kennicutt law (Kennicutt et al. 1998). 
In this analysis we have included only star-forming galaxies. 
To exclude passive sources 
we applied an empirical color selection of $(U-B)_{rf}<1.1$, calibrated from our data, and, to recover massive dusty sources 
that might fall into the red sequence, we included in the stacking analysis also sources with $(U-B)_{rf}>1.1$ and mag[24]-mag[3.6]$>$0.5.

\begin{figure}
   \centering
\includegraphics[width=9.cm]{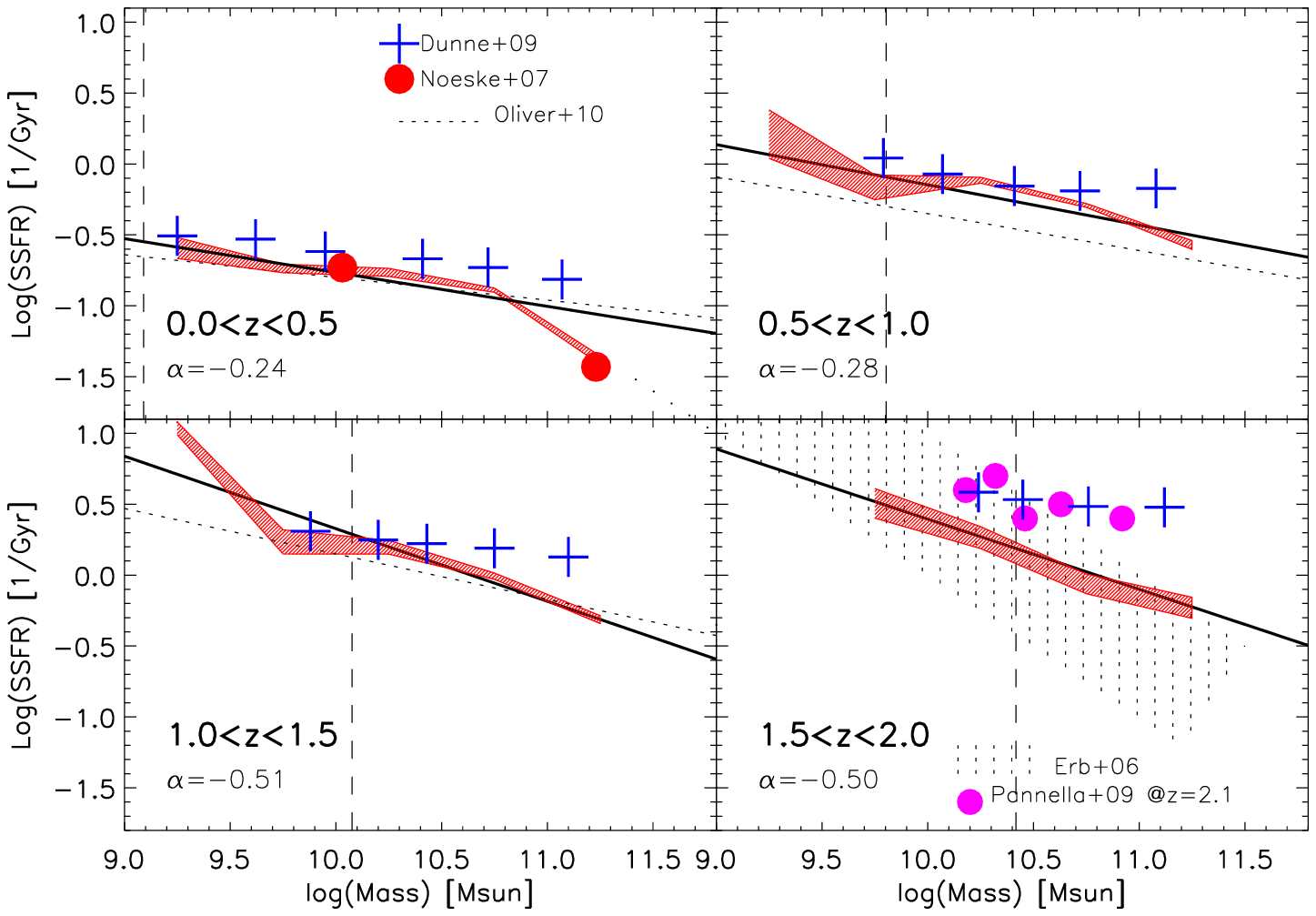}
\includegraphics[width=7.5cm]{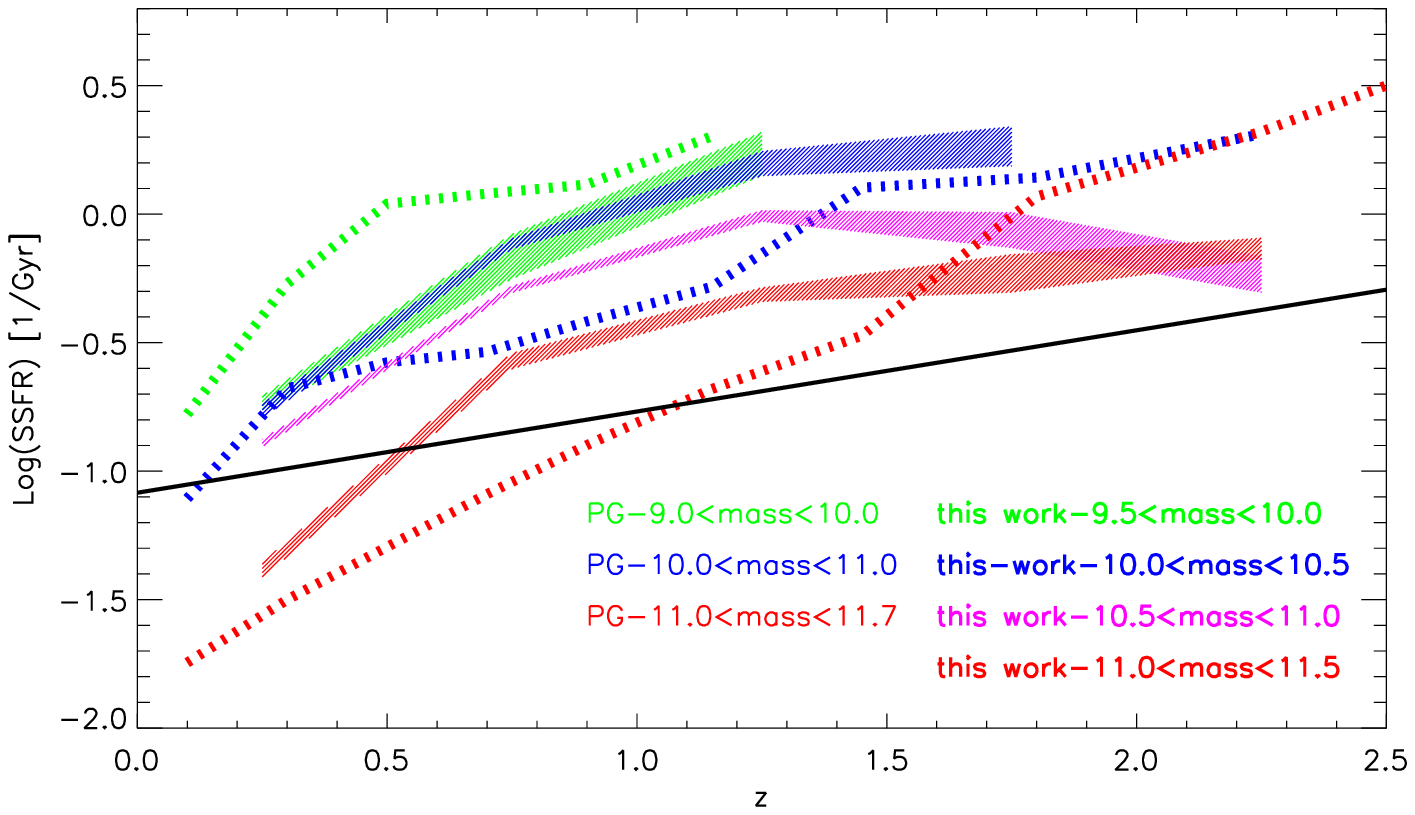}
\caption{\textit{Upper figure}: SSFR-mass relation based on a stacking analysis on PACS 
images for a sample of IRAC mag[4.5]$<$23.00 star-forming sources in 
various redshift bins. Our data are shown as a red shaded region, where 
the upper and lower envelops of each curve represent the statistical errors 
derived from a bootstrapping procedure on the stacking analysis. The black 
solid lines mark linear fits to our distribution. Literature data are 
reported with different symbols (see legend in the Figure). Vertical lines mark the mass completeness.
\textit{Lower figure}: SSFR-$z$ relation, in various mass bins. Our data (shaded regions) are compared to Perez-Gonzalez et al. (2008, dotted lines).
The black line is the inverse of time since Bing Bang at any redshift.
 }
\label{stacking}%
\end{figure}

The results of the PACS stacking analysis are presented in Figure \ref{stacking} (upper panel)  and compared with literature data based on star-forming samples (with
the exception of Dunne et al. 2009 that includes all $K$-band selected objects). 
The slope of our SSFR-mass relation becomes steeper with redshift, being quite flat
at $z<1$  with $\alpha=-0.25^{+0.11}_{-0.14}$ (all the slopes have been computed in a mass complete range).
Intringuingly, the slopes are consistent with those observed for the 24, 100 and 160$\mu$m detections at $0.5<z<2$ (blue, long-dashed lines in Figure 3). 
At $z<1$, our results are in broad agreement with those based on radio-stacking that found almost flat relations up to $z\sim2$ (Dunne et al. 2009, 
Pannella et a. 2009) , while at $z>1$ our relation evolves toward stronger dependencies ($\alpha=-0.50^{+0.13}_{-0.16}$).
We note that the rejection of passive and massive sources in the Dunne et al. sample would produce even flatter relations, increasing the discrepancy with our results at high-$z$.
At low $z$, the study by Noeske et al. (2007)  based on the UV-optical luminosities is steeper then our results, while at $z\sim2$ the work by Erb et al. (2006) 
is more consistent with our observed SSFR-mass relation (however, within a very huge scatter). 
A recent work by Oliver et al. (2010), based on MIPS 70 and 160$\mu$m stacking, seems instead to be in general agreement  with us up to 
$z<1.5$, where they have enough statistics.

Figure 4 (bottom panel) also shows that the mean SSFR of star-forming sources rises with redshift, 
up to a factor $\sim15$ for the most massive galaxies ($M>10^{11}M_{\odot}$), implying that galaxies tend to 
form their stars more actively at higher redshifts. The mean SSFR  seems also to flatten at $z>1.5$ for ($M>10^{10.5}M_{\odot}$).
Moreover, the most massive galaxies have the lowest SSFR at any redshifts. As shown in bottom Fig.4, they have already so large stellar masses at 
$z$=0.7 to 2.5 that they would require steady SFR at the PACS observed level during the whole Hubble time at that redshift to form.

\begin{acknowledgements}
GR acknowledges support from the University of Padova from ASI (Herschel Science Contract  I/005/07/0).         
PACS has been developed by a consortium of institutes led by MPE (Germany) and including UVIE 
(Austria); KU Leuven, CSL, IMEC (Belgium); CEA, LAM (France); MPIA (Germany); INAF- 
IFSI/OAA/OAP/OAT, LENS, SISSA (Italy); IAC (Spain). This development has been supported by the 
funding agencies BMVIT (Austria), ESA-PRODEX (Belgium), CEA/CNES (France), DLR (Germany), 
ASI/INAF (Italy), and CICYT/MCYT (Spain).
   \end{acknowledgements}

\Online

\begin{appendix}
\section{Authors affiliations}\label{sect:affiliations}
\begin{enumerate}[label=$^{\arabic{*}}$]
\item   Department of Astronomy, University of Padova, Vicolo dell'Osservatorio 3, I-35122 Padova, Italy.
              \email{giulia.rodighiero@unipd.it}
\item    Dipartimento di Astronomia, Universit\`a di Bologna, via Ranzani 1, I-40127 Bologna, Italy
\item    INAF-Osservatorio Astronomico di Bologna, Via Ranzani 1, I-40127, Bologna, Italy
\item    Max-Planck-Institut f\"{u}r  extraterrestrische Physik, Postfach 1312, 85741 Garching, Germany
\item   European Southern Observatory, Karl-Schwarzschild-Str. 2, 85748 Garching, Germany
\item  INAF-Osservatorio Astronomico di Roma, via di Frascati 33, 00040 Monte Porzio Catone, Italy.
\item  ESAC, Villafranca del Castillo, ES-28691 Madrid, Spain
\item  Laboratoire AIM, CEA/DSM-CNRS-Universit{\'e} Paris Diderot, IRFU/Service d'Astrophysique, B\^at.709, CEA-Saclay, 91191 Gif-sur-Yvette Cedex, France.
\item  Instituto de Astrof{\'i}sica de Canarias, 38205 La Laguna, Spain.
\item   Department of Astronomy, 610 Space Sciences Building, Cornell University, Ithaca, NY 14853, USA
\item  INAF-IFSI, Via Fosso del Cavaliere 100, I-00133 Roma, Italy
\item   INAF-Osservatorio Astronomico di Padova, Vicolo dell'Osservatorio 2, I-35122 Padova, Italy
\item   INAF-Osservatorio Astronomico di Trieste, via Tiepolo 11, 34143 Trieste, Italy
          
\end{enumerate}
\end{appendix}


\begin{appendix}
\section{Mass completeness}\label{mc}
\begin{figure}[ht!]
\centering
\includegraphics[width=9.5cm]{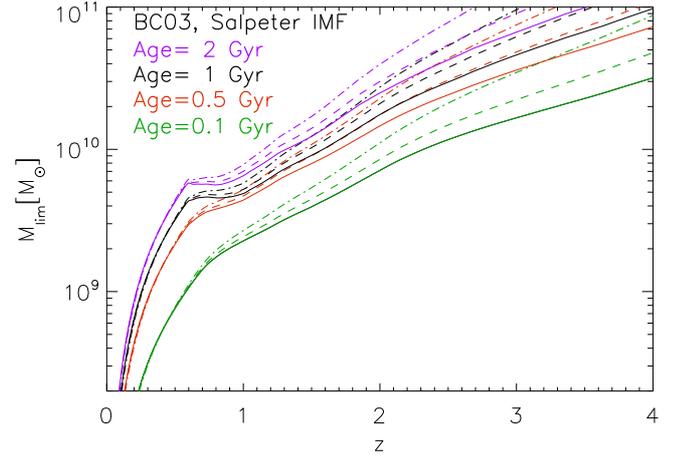}
\caption{Mass-completeness thresholds as a function of redshift for our IRAC 4.5$\mu$m-selected sample (mag$_{4.5 \mu m}<23.0$, AB), derived from synthetic stellar population models as described in Mancini et al. (2009). Here we used the constant SFR templates of Bruzual \& Charlot (2003), with a Salpeter IMF, and different ages, and dust extinction parameters ($E_{B-V}$). For each specific age, we considered three possible values of dust extinction: $E_{B-V}=0.3$ (solid lines), $E_{B-V}=0.5$ (dashed lines), and $E_{B-V}=0.8$ (dot-dashed lines). In our analysis we adopted the most conservative mass-completeness limit (dot-dashed magenta line), above which even the oldest (2 Gyr) and highly extincted ($E_{B-V}=0.8$) star-forming galaxy population would be entirely recovered.}
   \label{chiara}%
\end{figure}
\end{appendix}

\begin{appendix}
\section{SED fitting examples}\label{sedfit}

\begin{figure*}
\centering
\includegraphics{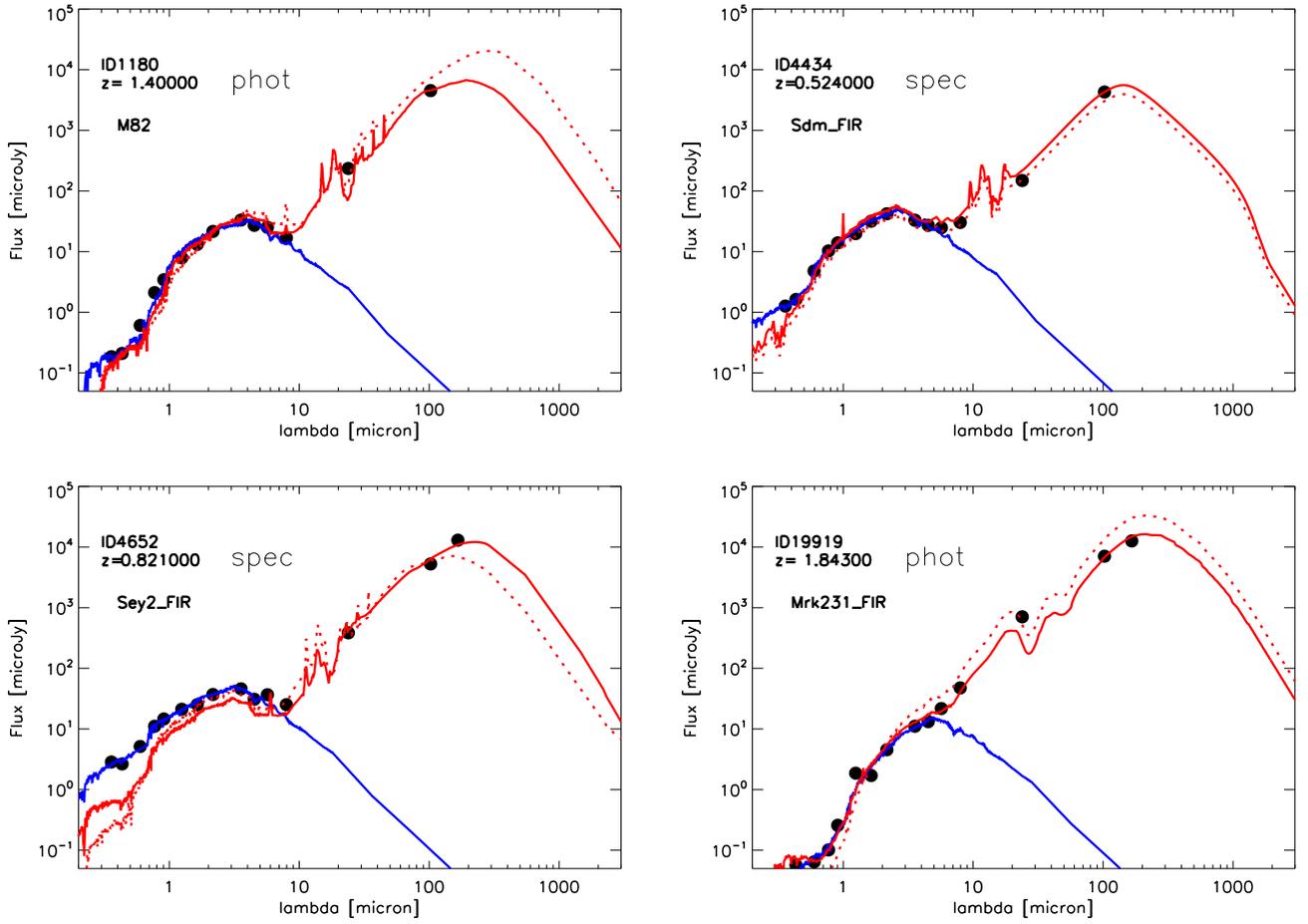}
 \caption{The observed Spectral Energy Distributions (filled circles) of four PACS sources from our sample, representative of the various spectral classes considered in this work. 
 We also show the best-fit spectra obtained with \textit{Hyperz} based on the spectral library by Polletta et al. (2007) and the updated version of Gruppioni et al. (2010): the red solid 
 lines represent the best-fit obtained with \textit{Herschel}  data, while red dotted lines are the fit limited to the 24$\mu$m \textit{Spitzer} data points.
 The blue lines show the fit to the optical-5.8 $\mu$m bands with Bruzual \& Charlot (2003) models . }
  \label{sedfitfig}%
\end{figure*}

\end{appendix}
\end{document}